\documentclass[12pt]{article}
\usepackage{cite,pifont,times} 
\usepackage{amssymb,amsbsy,amsmath,amscd,tikz}
\usepackage{wasysym}

\newcommand{\nnn}{\mathrm{NNN}}
\newcommand{\bcdot}{\boldsymbol{\cdot}}
\newcommand\email[4]{#1@#2.#3.#4}
\def\C{{\mathbf{C}}}
\def\N{\mathcal{N}}
\def\L{{\Lambda}}
\def\R{{\mathbf{R}}}
\def\t{\theta}
\newcommand\G[1]{{\Gamma\left(#1\right)}}
\def\eq#1{(\ref{#1})}

\newcommand\laplacian{\bigtriangleup}

\newcommand\bcc{
\tikz\draw[scale=.14] (0,0,0) -- (1,1,1) -- (-1,1,1) -- (-1,-1,1) -- (1,-1,1) -- (1,1,1)
--(1,1,-1) -- (-1,1,-1) -- (-1,-1,-1) -- (1,-1,-1) -- (1,1,-1)
(0,0,0) -- (1,-1,1) -- (1,-1,-1) -- (0,0,0) -- (-1,-1,1) -- (-1,-1,-1) --
(0,0,0) -- (-1,1,1) -- (-1,1,-1) -- (0,0,0);
}
\begin{document}
\title{Green's function on  lattices}
\author{ 
Koushik Ray \thanks{\email{koushik}{iacs}{res}{in}} \\
\small Department of Theoretical Physics,
\small  Indian Association for the Cultivation of Science \\
\small  Calcutta 700~032. India.
}
\date{}
\maketitle
\begin{abstract}
\noindent 
A method to calculate exact Green's functions on lattices in 
various dimensions is presented. Expressions in terms of generalized 
hypergeometric functions in one or more variables are obtained for various
examples by relating the resolvent to a contour integral,
evaluated using residues. Different ways of arranging the series
leads to different combinations of hypergeometric functions 
providing identities involving generalized hypergeometric functions.
The method is shown to be useful for computing Green's functions 
with next-nearest neighbor hopping as well.
\end{abstract}
\thispagestyle{empty}
\section{Introduction}
\noindent 
A variety of physical situations call for studying Green's function on
lattices. Examples include crystal Physics \cite{montroll}, 
electrical circuits \cite{circuit1,circuit2}, statistical Physics 
\cite{ID,priz1,priz3}, lattice gauge theory
\cite{necco,weisz} etc. to mention a few. 
Various aspects of lattice Green's functions have been
studied on a variety of lattices in diverse dimensions
\cite{longer,joyce1,berciu,binom,guttmann,maass,hollos}.

In this article we present a method to compute the exact form of Green's 
functions on lattices with a few examples.   
Given a lattice, the Green's function, defined as the kernel of the
Laplacian on the same is first expressed  as  Fourier integrals 
over closed intervals, as many as the dimension of the lattice in number. 
Each of these closed domains 
is expressed in turn as the unit circle in an appropriately defined complex 
plane, promoting the Fourier integral to a contour integral. The latter can
be evaluated by the method of residues. The Green's function thus obtained is,
more often than not, singular. We resort to evaluating the resolvent 
of the Laplacian  which is, by the same token, a contour integral containing 
an auxiliary complex parameter, the spectral parameter. 
Computation of residue yields a series in the spectral parameter, 
whose domain is restricted by the radius of  convergence of the series. 
It is usually possible 
to analytically continue the result to ulterior domains, appropriate to the
specificity of physical situations. Indeed, in most cases we are able to 
express the resolvent in terms of generalized hypergeometric functions 
for which rigorous results of analytic continuation are available
\cite{buh,skor,NIST}.

In the following sections we study the Laplacian on lattices in various 
dimensions. In the rest of this section let us recall the relevant 
definitions in order to fix notation. 
Let $\L$ denote an infinitely extended lattice in the $D$-dimensional 
Euclidean space $\R^D$, with a point marked as the origin. 
Let $\N$ denote the set of points on the lattice adjacent to the origin, that
is the set of nearest neighbors of the origin, according to the canonical
Euclidean metric on $\R^D$. 
Let $N$ denote the the cardinality of $\N$, that is the
number of nearest neighbors to any point of $\L$, called the co-ordination
number of the lattice. For any element $a$ of $\N$, let $\nabla_a$ denote 
the difference operator on the space of complex-valued functions in 
$\R^D$, that is
\begin{equation} 
\nabla_a f(\xi) = f(\xi+a)-f(\xi), 
\end{equation} 
where $f:\R^D\to\C$ and $\xi\equiv (\xi_1,\cdots, \xi_D)\in\R^D$.
The second order difference operator on $\L$ is then
\begin{equation}
\label{delta}
\begin{split}
\nabla^2 f(\xi) &= \sum\limits_{a,b\in\N} \nabla_a\nabla_b f(\xi)\\
&=\sum_{a,b\in\N} f(\xi+a+b) -2N\sum_{a\in\N} f(\xi+a) + N^2f(\xi).
\end{split}
\end{equation} 
Shifting $\xi$ to $\xi-a$ and dividing by $N$ leads to the Laplacian 
$\laplacian$ defined as 
\begin{equation}
\label{laplace}
\laplacian f(x) = \frac{1}{2}\sum_{a\in\N}\big( f(\xi+a) + f(\xi-a) 
- 2f(\xi)\big).
\end{equation} 
The normalizing factor of half is introduced to match with the 
standard expression on lattices for which both $a$ and $-a$ belong to $\N$. 
The Green's function $G_a$  is then obtained as the 
kernel of the Laplacian $\laplacian$, so that
\begin{equation}
\laplacian G(\xi) = -\delta(\xi).
\end{equation} 
In order to find the Green's function let us write any complex-valued
function$f(x)$ as a Fourier integral,
\begin{equation}
\label{fourier}
f(\xi) = \frac{1}{(2\pi)^D} \int_0^{2\pi}d^D\t e^{i\xi\bcdot\t} g(\t),
\end{equation} 
where $\t\in\R^D$ and $\bcdot$ denotes the dot product in $\R^D$.
Substituting in \eq{laplace} we obtain 
\begin{equation}
\laplacian f(\xi)=\frac{1}{(2\pi)^D}\int_0^{2\pi}d^D\t e^{i\xi\bcdot\t}\ 
g(\t) \Big(\frac{1}{2}\sum_{a\in\N}\big(
e^{ia\bcdot\t} + e^{-ia\bcdot\t} \big)-N\Big), 
\end{equation} 
which reduces to $-\delta(\xi)$ if and only if 
\begin{equation}
g(\t) =  \frac{1}{N-\frac{1}{2}\sum\limits_{a\in\N}\big(e^{ia\bcdot\t} 
+ e^{-ia\bcdot\t} \big) }.
\end{equation} 
The two-point correlation function for a pair of lattice
points $\alpha$, $\beta$ is defined as 
\begin{equation}
\label{green}
{\mathcal G}(\alpha,\beta) = G(\alpha,\beta) - G(0,0).
\end{equation} 
Subtraction of $G(0,0)$ is to impose the translational symmetry of the
lattice $\L$, required since we started with the origin marked in $\L$. 
Introducing the vector $r=\alpha-\beta$ with integer components when expanded
in the basis of the lattice vectors, 
$r = (r_1,r_2,\cdots, r_D)$ we write
\begin{equation}
\label{green1}
\begin{split}
G_{r_1,r_2,\cdots,r_D} 
=G(\alpha,\beta)
&= \frac{1}{(2\pi)^D}\int_0^{2\pi} d^D\t \frac{e^{ir\bcdot\t}}{%
N-\frac{1}{2}\sum\limits_{a\in\N} \big(e^{ia\bcdot\t} 
+ e^{-ia\bcdot\t} \big)},\\
&= \frac{1}{(2\pi)^D}\int_0^{2\pi} d^D\t \frac{e^{ir\bcdot\t}}{%
N-\sum\limits_{a\in\N} \cos a\bcdot\t}.
\end{split}
\end{equation}  
The integral as defined is usually singular. We recourse to computing the
resolvent 
\begin{equation} 
\label{reslv}
H_r(t)= \frac{1}{(2\pi)^D}\int_0^{2\pi} d^D\t \frac{e^{ir\bcdot\t}}{%
t-\sum\limits_{a\in\N} \cos a\bcdot\t},
\end{equation} 
where $t$ is a complex parameter, the spectral parameter. By abuse of notation 
we shall continue denoting the resolvent by $H(t)$ even though 
$t$ is normalized differently in different examples for convenience.
The general strategy adopted here to evaluate the resolvent 
is to first introduce a complex variable for every 
component of $\t$. The integral over the angular variables are then 
interpreted as a contour integral over the unit circle in each of these 
complex planes. The contour integral is then evaluated using the method 
of residues \cite{senray}. Let us now employ the strategy in a 
variety of examples in various dimensions. 
\section{One dimension}
The set of nearest neighbors to the origin in a one-dimensional lattice
is $\N=\{-1,1\}$. The co-ordination number is $N=2$. 
The Green's function is obtained from \eq{green1}  to be 
\begin{equation}
G^{(1)}_{r} = \frac{1}{2\pi} \int_0^{2\pi}d\t
\frac{e^{ir\t}}{2-2\cos\t }.
\end{equation} 
The resolvent \eq{reslv} is written as a contour integral in terms of a
complex variable $x=e^{i\t}$ as
\begin{equation}
H_r(t) = \frac{1}{2\pi i}\oint\limits_{|x|=1}\frac{dx}{x}\ 
\frac{x^r}{2t-(x+1/x)}.
\end{equation} 
The Green's function is $G^{(1)}_{r} = H_{r}(1)$. 
In order to evaluate the contour integral we rewrite it by first 
pulling out $2t$ from the denominator and then expanding the denominator 
in a geometric series to obtain
\begin{equation} 
\begin{split}
H_r(t) &= \frac{1}{2\pi i}\oint\limits_{|x|=1}\frac{dx}{x}\ 
{x^r} \sum_{n=0}^{\infty} (2t)^{-1-n} (x+1/x)^n\\
&= \frac{1}{2\pi i}\oint\limits_{|x|=1}\frac{dx}{x} \sum_{n=0}^{\infty}
(2t)^{-1-n} \sum_{a=0}^n \binom{n}{a} x^{2a-n+r},
\end{split}
\end{equation} 
where we have used the binomial theorem to expand $(x+1/x)^n$ in the second
step. 
Writing the factorials in the binomial coefficients as gamma functions the
integral takes the form 
\begin{equation} 
H_{r,s}(t) = 
\frac{1}{2\pi i}\oint\limits_{|x|=1}\frac{dx}{x} \sum_{n=0}^{\infty}
\sum_{a=0}^{\infty} (2t)^{-1-n} \frac{\G{1+n}}{\G{1+n-a}\G{1+a}}\
x^{2a-n+r},
\end{equation} 
where the domain of the $a$ is extended taking into account the poles of
gamma function for non-positive integers appearing in the denominator.
The integral is then evaluated as the residue of the integrand.
Since the measure is $dx/x$, the residue is given by the constant, that is,
the totality of  $x$-independent terms of the integrand, with
\begin{equation}
\label{Hcons}
2a-n+r=0.
\end{equation} 
This equation can be solved for $a$ as $a = (n-r)/2$, leading to
\begin{equation}
H_r(t) = \sum_{n=0}^{\infty} \left(\frac{1}{2t}\right)^{1+n} 
\frac{\G{1+n}}{\G{1+\frac{n+r}{2}}\G{1+\frac{n-r}{2}}}.
\end{equation} 
Splitting the sum over all positive integers into even and odd parts as
\begin{equation}
\label{splitsum}
\sum_{n=0}^{\infty} f(n) = 
\sum_{n=0}^{\infty} f(2n) + 
\sum_{n=0}^{\infty} f(2n+1)  
\end{equation} 
for any function $f$ and using the duplication formula 
\begin{equation} 
\label{dup}
\G{2x}=\frac{2^{2x-1}}{\sqrt{\pi}}\G{x}\G{x+{1}/{2}},
\end{equation} 
the sum is written as
\begin{equation}
H_r(t) = \frac{1}{2\sqrt{\pi}t} 
\sum_{n=0}^{\infty} \left(\frac{1}{t^2}\right)^{n} 
\frac{\G{\frac{1}{2}+n}\G{1+n}}{\G{\frac{2+r}{2}+n}\G{\frac{2-r}{2}+n}}
+ \frac{1}{2\sqrt{\pi}t^2} 
\sum_{n=0}^{\infty} \left(\frac{1}{t^2}\right)^{n} 
\frac{\G{\frac{3}{2}+n}\G{1+n}}{\G{\frac{3+r}{2}+n}\G{\frac{3-r}{2}+n}}.
\end{equation} 
This can be expressed in terms of hypergeometric functions as 
\begin{equation}
\label{1d:hyp}
H_r(t) = \frac{1}{2\sqrt{\pi}t} 
\frac{\G{1}^2\G{\frac{1}{2}}}{\G{\frac{2+r}{2}}\G{\frac{2-r}{2}}}
{}_3F_2\left(\left.%
\genfrac{}{}{0pt}{}{1,1,\frac{1}{2}}{\frac{2+r}{2},\frac{2-r}{2}}
\right|\frac{1}{t^2}\right)
+ \frac{1}{2\sqrt{\pi}t^2} 
\frac{\G{1}^2\G{\frac{3}{2}}}{\G{\frac{3+r}{2}}\G{\frac{3-r}{2}}}
{}_3F_{2}\left(\left.
\genfrac{}{}{0pt}{}{1,1,\frac{3}{2}}{\frac{3+r}{2},\frac{3-r}{2}}
\right|\frac{1}{t^2}\right).
\end{equation} 
Using the relations
$\G{1-z}\G{z}=\pi/\sin\pi z$, $\G{z+1}=z\G{z}$, for $\mathop{Re} z> 0$,
the values 
$\G{1/2} = \sqrt{\pi}$, $\G{1}=1$ leading to the expressions
\begin{gather}
\G{1+\frac{r}{2}}\G{1-\frac{r}{2}} = \frac{\pi r}{2\sin\frac{\pi r}{2}},\\
\G{\frac{3+r}{2}}\G{\frac{3-r}{2}} = 
\frac{\pi (1-r^2)}{4\cos\frac{\pi r}{2} },
\end{gather} 
we finally arrive at 
\begin{equation} 
\label{HD1:1}
H_r(t) = \frac{1}{\pi r t}\ {\sin \left(\frac{\pi r}{2}\right)}
{}_3F_{2}\left(\left.
\genfrac{}{}{0pt}{}{1,1,\frac{1}{2}}{\frac{2+r}{2},\frac{2-r}{2}}
\right|\frac{1}{t^2}\right) 
+ \frac{1}{\pi (1-r^2) t^2}\ {\cos \left(\frac{\pi r}{2}\right)}
{}_3F_{2}\left(\left.
\genfrac{}{}{0pt}{}{1,1,\frac{3}{2}}{\frac{3+r}{2},\frac{3-r}{2}}
\right|\frac{1}{t^2}\right).
\end{equation} 
The Green's function is given by the value of $H_r$ at $t=1$. 

The expression is manifestly symmetric under the reflection $r\to -r$,
which is a symmetry of the one-dimensional lattice. It is not best-suited 
for analyzing the asymptotics, however. The appearence of gamma functions
with both signs of $r$ in their arguments makes it difficult to  take limits
of $r$. In order to obtain a formula better adapted
for such studies, let us solve \eq{Hcons} as  
\begin{equation}
n=r+2a, \quad r\geqslant 0,
\end{equation} 
which leads to the series
\begin{equation}
\label{1d:ser}
H_r(t)=\sum_{a=0}^{\infty} \left(\frac{1}{2t}\right)^{1+r+2a} 
\frac{\G{2a+1+r}}{\G{1+a}\G{1+r+a}}.
\end{equation} 
Using the duplication formula \eq{dup} in the numerator we obtain
\begin{equation}
\label{HD1:2}
H_r(t) = \left(\frac{1}{2t}\right)^{1+r} {}_2F_{1}\left(\left.
\genfrac{}{}{0pt}{}{1+\frac{r}{2},\frac{1+r}{2}}{1+r}\right|\frac{1}{t^2}
\right),
\end{equation} 
or, equivalently \cite[{\small\bfseries 15.4.18}]{NIST},
\begin{equation}
H_r(t) = \frac{1}{2\sqrt{t^2-1}}\frac{1}{(t+\sqrt{t^2-1})^r}. 
\end{equation} 
This reproduces the well-know formula for $r=0$.
Equating the expressions \eq{HD1:1} and \eq{HD1:2} furnishes an identity of
hypegeometric functions. Asymptotic behavior of $H$ and the Green's 
function can be studied using existing results \cite{paris,temme,NIST}.
\section{Two dimensions}
\label{D:2}
There are various kinds of lattices in two dimensions. We consider
Green's function on three such in this section. 
Using existing results on analytic continuation of
hypergeometric functions we demostrate the cancellation of singularity in
\eq{green} for the square lattice. 
\subsection{Square lattice}
\label{sq:2}
For the square lattice, the set of nearest neighbors is 
$\N=\{(\pm 1,0),(0,\pm 1)\}$ with cardinality $N=4$. 
The Green's function is 
\begin{equation}
G^{\Square}_{p,q}=\frac{1}{4\pi^2}
\int\limits_{0}^{2\pi} 
\int\limits_{0}^{2\pi} 
\frac{e^{ip\t_1+iq\t_2}\ d\t_1\ d\t_2}{4-2\cos\t_1-2\cos\t_2}.
\end{equation} 
Isotropy of the square lattice requires the Green's function to satisfy
\begin{equation}
\label{Gsymm}
\mathcal{G}^{\Square}_{q,p} 
=\mathcal{G}^{\Square}_{-p,q} 
=\mathcal{G}^{\Square}_{p,-q} 
=\mathcal{G}^{\Square}_{p,q}
\end{equation} 
We deal with the resolvent written as a contour integral 
\begin{equation}
H_{r,s}(t)=
\frac{1}{(2\pi i)^2}\oint\limits_{|x|=1}\oint\limits_{|y|=1} 
\frac{dx}{x}\frac{dy}{y}\
\frac{x^r\ y^s}{4t-(x+1/x)(y+1/y)}
\end{equation} 
in two complex variables $x$ and $y$, where the contour is the union of unit
circles in each complex plane. Parametrizing the circles as phases with 
$x=e^{i(\t_1+\t_2)/2}$ and $y=e^{i(\t_1-\t_2)/2}$ relates the resolvent and
the Green's function through
\begin{equation}
\label{H2G}
G^{\Square}_{p,q} = \frac{1}{8\pi^2}\ H_{p+q,p-q}(1).
\end{equation} 
The integral $H_{r,s}(t)$ is  evaluated by computing the residue of the
integrand at the poles inside the unit circles. By first pulling
out $4t$ from the denominator and then expanding the denominator in a
binomial series we have
\begin{equation}
\begin{split}
H_{r,s}(t) 
=\sum_{n=0}^{\infty} (4t)^{-1-n}\ 
\oint\limits_{|x|=1}\oint\limits_{|y|=1}\frac{dx}{x}\frac{dy}{y}\
(x+1/x)^n(y+1/y)^n x^ry^s. 
\end{split}
\end{equation} 
Expanding the factors $(x+1/x)^n$ and 
$(y+1/y)^n$ in binomial series then leads to  
\begin{equation}
\begin{split}
H_{r,s}(t) 
=
\oint\limits_{|x|=1}\oint\limits_{|y|=1}\frac{dx}{x}\frac{dy}{y}\
\sum_{n=0}^{\infty} (4t)^{-1-n}\ 
\sum_{a,b=0}^{n}\
\binom{n}{a}\binom{n}{b} x^{r+2a-n}y^{s+2b-n}. 
\end{split}
\end{equation} 
Due to the presence of inverses of $x$ and $y$ in the measure the residue 
of the integrand is given by the terms of the sum which are independent 
of $x$ and $y$, as before. These constant terms lead to the resolvent 
\begin{equation}
H_{r,s}(t) =\sum_{n,a,b=0}^{\infty} \left(\frac{1}{4t}\right)^{1+n} 
\frac{\G{1+n}^2}{\G{1+a}\G{1+b}\G{1+n-a}\G{1+n-b}},
\end{equation}  
with the indices satisfying the constraints 
\begin{equation}
\label{sq:ind}
\begin{split}
r+2a-n &=0,\\
s+2b-n &=0.
\end{split} 
\end{equation}
The domain of the indices $a$ and $b$ are extended as before since gamma
functions having poles for non-positive integers appear in the denominator
for the extra terms.
Solving these for $a$ and $b$ with $a=(n-r)/2$ and $b=(n-s)/2$ yields  
\begin{equation} 
\label{hrs:Gm}
H_{r,s}(t) =\sum_{n=0}^{\infty} \left(\frac{1}{4t}\right)^{1+n} 
\frac{\G{1+n}^2}{\G{1+\frac{n+r}{2}} \G{1+\frac{n-r}{2}}
\G{1+\frac{n+s}{2}} \G{1+\frac{n-s}{2}}}.
\end{equation} 
Splitting the sum over all positive integers into even and odd parts 
as in \eq{splitsum}
and using the duplication formula \eq{dup} as in the previous section,
the above expression is written in terms of the zero-balanced
generalized hypergeometric function ${}_5F_4$ as
\begin{equation}
\label{hrs:eq}
\begin{split}
H_{r,s}(t) &=
\frac{1}{4\pi t}\
\frac{\G{1}^3\G{\frac{1}{2}}^2}{%
\G{1+\frac{r}{2}}\G{1-\frac{r}{2}}\G{1+\frac{s}{2}}\G{1-\frac{s}{2}}}\ %
{}_5F_{4}\left(\left.
\genfrac{}{}{0pt}{}{1,1,1,\frac{1}{2},\frac{1}{2}}{
\frac{2+r}{2},\frac{2-r}{2},\frac{2+s}{2},\frac{2-s}{2} 
}
\right| \frac{1}{t^2} \right) \\
&\quad+ 
\frac{1}{4\pi t^2}\
\frac{\G{1}^3\G{\frac{3}{2}}^2}{%
\G{\frac{3+r}{2}}\G{\frac{3-r}{2}}\G{\frac{3+s}{2}}\G{\frac{3-s}{2}}}\ %
{}_5F_{4}\left(\left.
\genfrac{}{}{0pt}{}{1,1,1,\frac{3}{2},\frac{3}{2}}{
\frac{3+r}{2},\frac{3-r}{2},\frac{3+s}{2},\frac{3-s}{2} 
}
\right| \frac{1}{t^2} \right)
\end{split} 
\end{equation} 
in the domain $|t|>1$.
Substituting $r=p+q$ and $s=p-q$ we obtain 
\begin{equation}
\begin{split}
H_{p+q,p-q}(t) &= 
\frac{1}{2\pi^2}\frac{1}{t}\
\frac{\cos(\pi p)-\cos(\pi q)}{p^2-q^2}\ 
{}_5F_{4}\left(\left.
\genfrac{}{}{0pt}{}{1,1,1,\frac{1}{2},\frac{1}{2}}{
\frac{2+p+q}{2},\frac{2-p-q}{2},\frac{2+p-q}{2},\frac{2-p+q}{2} 
}
\right|\frac{1}{t^2}\right) \\
&\quad+ 
\frac{1}{2\pi^2}\frac{1}{t^2}\
\frac{\cos(\pi p)+\cos(\pi q)}{((p+q)^2-1)((p-q)^2-1)}\
{}_5F_{4}\left(\left.
\genfrac{}{}{0pt}{}{1,1,1,\frac{3}{2},\frac{3}{2}}{
\frac{3+p+q}{2},\frac{3-p-q}{2},\frac{3+p-q}{2},\frac{3-p+q}{2} 
}
\right|\frac{1}{t^2}\right). 
\end{split}
\end{equation} 
This gives the Green's function $G_{p,q}$ by \eq{H2G},
possessing the symmetries \eq{Gsymm}, which is singular as the
hypergeometric function is singular at $t=1$.
The singularity is cancelled by subtracting $G_{0,0}$. To see this we first 
analytically continue $H_{r,s}(t)$ to a neighborhood of $t=1$ to isolate the
logarithmic singularity. We then verify that the coefficient of the
logarithmic term is independent of $p,q$. 
The expression \eq{hrs:eq} continues to \cite{buh}
\begin{equation} 
\label{hrs:1}
\begin{split}
H_{r,s}(t) &=
\sum_{n=0}^{\infty} (1-\frac{1}{t^2})^n
\left[
\sum_{k=0}^n\frac{(-n)_k}{k!k!} 
\Big(\frac{A^{(4)}_k}{4\pi t} + \frac{A^{\prime(4)}_k}{4\pi t^2}\Big)
\big(\psi(1+n-k)-\psi(1+n)-\log(1-\frac{1}{t^2})\big)\right.\\
&\left.\qquad+
(-1)^n n! 
\sum_{k=n+1}^{\infty} \frac{(k-n-1)!}{k!k!}
\Big(\frac{A^{(4)}_k}{4\pi t} + \frac{A^{\prime(4)}_k}{4\pi t^2}\Big)
\right], 
\end{split}
\end{equation} 
in a neighborhood of $t=1$, where we used the Pocchammer symbol
\begin{equation}
(\ell)_n = \ell(\ell+1) \cdots(\ell+n-1)={\G{\ell+n}}/{\G{\ell}},
\end{equation} 
which for positive integral $n$ satisfies  $(1)_n=n!$, 
$(\ell)_0=1$ for all $\ell$, $(0)_0=1$, in particular, and
$(0)_n=0$ for non-vanishing $n$. 
The dependence on $r$ and $s$ are solely in the coefficients
$A$ and $A'$, defined as \cite{buh}
\begin{equation} 
\begin{split}
A^{(4)}_k &= \frac{(1+\frac{r}{2})_k(1-\frac{r}{2})_k}{k!}
\sum_{\ell=0}^{\infty} \frac{(\frac{1}{2})_{\ell}(1)_{\ell}}{
\ell! (1+\frac{r}{2})_{\ell}(1-\frac{r}{2})_{\ell}}
{}_3F_2\left(
\left.
\genfrac{}{}{0pt}{}{\frac{1+s}{2},-\frac{1+s}{2},-\ell}{\frac{1}{2},1}
\right|1 \right),
\\
A^{\prime(4)}_k &= \frac{(1+\frac{r}{2})_k(-\frac{1+r}{2})_k}{k!}
\sum_{\ell=0}^{\infty} \frac{(\frac{1}{2})_{\ell}(0)_{\ell}(-k)_{\ell}}{
\ell! (1+\frac{r}{2})_{\ell}(-\frac{1+r}{2})_{\ell}}
{}_3F_2\left(
\left.
\genfrac{}{}{0pt}{}{\frac{s}{2},-\frac{s}{2},-\ell}{\frac{1}{2},0}
\right|1 \right)
\end{split}
\end{equation} 
From \eq{hrs:1} we note that the logarithmic singularity of $H_{r,s}$ 
arises from the $n=0$ term in the sum. The coefficient of $\log(1-1/t^2)$,
namely, 
\begin{equation} 
- \frac{A^{(4)}_0}{4\pi t}- \frac{A^{\prime(4)}_0}{4\pi t^2},
\end{equation} 
evaluates to $-1/2\pi$ as $t$ tends to unity, 
independent of $r$ or $s$. Hence the Green's function 
\eq{green} defined as the limit
\begin{equation} 
\mathcal{G}^{\Square}_{p,q}= \lim_{t\to 1}
\frac{1}{8\pi^2} \big(H_{p+q,p-q}(t)-H_{0,0}(t)\big)
\end{equation} 
is non-singular.

As in the one-dimensional case, the expression \eq{hrs:eq}, while 
possessing the symmetries of the lattice, is not amenable to asymptotic 
studies. We can derive another expression for the resolvent by solving 
\eq{sq:ind} for $n$ as $n=2a+r$, leading to
\begin{equation}
b= a+\frac{r-s}{2},\quad n-b = a+\frac{r+s}{2}.
\end{equation} 
Substituting these in \eq{hrs:Gm} and using the duplication formula \eq{dup}
we obtain $H_{r,s}$ in terms of another zero-balanced generalized 
hypergeometric function 
\begin{align}
\label{2sq:2}
H_{r,s}(t) &= \left(\frac{1}{4t}\right)^{1+r} \sum\limits_{n=0}^{\infty}
\frac{\G{1+r+2n}^2}{\G{1+r+n}\G{1+\frac{r+s}{2}+n}\G{1+\frac{r-s}{2}+n}}
\left(\frac{1}{16t^2}\right)^n \frac{1}{n!}
\\
&= 
\frac{\G{1+r}}{\G{1+\frac{r+s}{2}}\G{1+\frac{r-s}{2}}} \
\left(\frac{1}{2t}\right)^{1+r} 
{}_4F_{3}\left(\left.
\genfrac{}{}{0pt}{}{1+\frac{r}{2},1+\frac{r}{2},\frac{1+r}{2}, 
\frac{1+r}{2}}{1+\frac{r+s}{2},1+\frac{r-s}{2},1+r}
\right|\frac{1}{t^2}\right).
\end{align}
Isotropy \eq{Gsymm} of the lattice is not manifest in the expression for the
resolvent $H_{p+q,p-q}(t)$. 
For $r=s=0$ this yields
\begin{equation}
H_{0,0}(t) =  \frac{1}{\pi t}K({1}/{t}),
\end{equation} 
where the identification  of the hypergeometric function and the complete
elliptic integral of the first kind, namely, 
${}_2F_1\left(\left.
\genfrac{}{}{0pt}{}{1/2,1/2}{1}\right|x\right)=\frac{2}{\pi} K(x)$, is used. 
For large distances, $G_{p,p}$ is obtained, according to \eq{H2G}, as
$H_{2p,0}(1)$ for $p\gg 1$. 
Equating the two expressions \eq{hrs:eq} and \eq{2sq:2} 
for the same resolvent furnishes
another identity for generalized hypergeometric functions. 
Asymptotic logarithmic behavior of the Green's function $\mathcal{G}_{p,q}$
has been studied in detail \cite{ID,scaling}.
%
%
\subsection{Triangular and honeycomb lattices}
Another class of two-dimensional lattices that we consider next are the
triangular and honeycomb lattices, which are closely related. Indeed, they
can be obtained from the same resolvent by changing the definition of the
angular variables. In order to compute the Green's function
on triangular and honeycomb lattices \cite{guttmann,triang,honey} 
let us start from the integral 
\begin{equation}
\label{Hboth}
H_{r,s}(t)= \frac{1}{(2\pi i)^2}
\oint\limits_{|x|=1}\oint\limits_{|y|=1} \frac{dx}{x}\frac{dy}{y}
\frac{x^ry^s}{t-(x+y+1/xy)(1/x+1/y+xy)}
\end{equation} 
over unit circles in the $x$- and $y$-planes.
For the honeycomb lattice with $\N = \{(1,0),(0,1),(-1,-1)\}$ and $N=3$. 
Choosing angular variables $\theta_1$ and $\theta_2$ such that 
\begin{equation}
e^{i\theta_1}=x/y,\ e^{i\theta_2}= xy^2,
\end{equation} 
we have $x^2y=e^{i(\theta_1+\theta_2)}$. Consequently, the Green's function
on a honeycomb lattice
\begin{equation} 
G^{\varhexagon}_{p,q} =\frac{1}{4\pi^2}\int_0^{2\pi}\int_0^{2\pi}
\frac{e^{ip\theta_1+iq\theta_2}d\theta_1
d\theta_2}{3-\big(\cos\theta_1+\cos\theta_2+\cos(\theta_1+\theta_2)\big)}
\end{equation} 
is written in terms of \eq{Hboth} as
\begin{equation} 
G^{\varhexagon}_{p,q} = -6 H_{p+q,2q-p}(9).
\end{equation} 
Similarly, for the triangular lattice 
with $\N=\{(\pm 2,0),(\pm 1,\pm 1)\}$ and $N=6$.
With the choice of angular variables as 
\begin{equation} 
x^2y=e^{2i\theta_1},\ xy^2=e^{i(\theta_1+\theta_2)},
\end{equation} 
we have $x/y=e^{i(\theta_1-\theta_2)}$.
The  Green's function on the triangular lattice 
\begin{equation} 
G^{\bigtriangleup}_{p,q} =\frac{1}{4\pi^2}\int_0^{2\pi}\int_0^{2\pi}
\frac{e^{ip\theta_1+iq\theta_2}d\theta_1
d\theta_2}{6-2\big(\cos\ 2\t_1+\cos (\t_1-\t_2)+\cos (\t_1+\t_2)\big)}
\end{equation} 
is related to \eq{Hboth} by 
\begin{equation} 
G^{\bigtriangleup}_{p,q} = -\frac{3}{2} H_{p,\frac{p+3q}{2}}(9).
\end{equation} 
It thus suffices to consider the integral \eq{Hboth} which is again evaluated
by first expanding the denominator as
\begin{equation}
\begin{split}
H_{r,s}(t) 
&=  (2\pi i)^2
\sum_{n=0}^{\infty} t^{-1-n} 
\oint\limits_{|x|=1}\oint\limits_{|y|=1} \frac{dx}{x}\frac{dy}{y}
x^ry^s(x+y+\frac{1}{xy})^n
(\frac{1}{x}+\frac{1}{y}+xy)^n\\
&=  (2\pi i)^2
\sum_{n=0}^{\infty} 
\sum_{{i,j,a,b=0}\atop i+j\leqslant n,a+b\leqslant n}^n t^{-1-n} 
\oint\limits_{|x|=1}\oint\limits_{|y|=1} \frac{dx}{x}\frac{dy}{y}
x^{r+2a+b-2i-j} y^{s+a+2b-i-2j} 
\end{split}
\end{equation} 
and evaluating residue by collecting the constant part of the integrand by
choosing
\begin{equation}
i=3a+(2r-s)/3,\  j=b+(2s-r)/3,
\end{equation} 
to derive
\begin{equation}
H_{r,s}(t)= \sum_{n=0}^{\infty}\sum_{{i,j=0}\atop i+j\leqslant n}^n
t^{-1-n}\frac{n!n!}{i!j!(n-i-j)!(i+\frac{s-2r}{3})!(j+\frac{r-2s}{3})!
(n-i-j+\frac{r+s}{3})!}. 
\end{equation} 
Inserting the appropriate values of $r$ and $s$ we then have,
\begin{equation}
H_{p+q,2q-p}(t)= \sum_{n=0}^{\infty}\sum_{{i,j=0}\atop i+j\leqslant n}^n
t^{-1-n}\frac{n!n!}{i!j!(n-i-j)!(i-p)!(j+p-q)!(n-i-j+q)!}
\end{equation} 
for the honeycomb lattice and 
\begin{equation}
H_{p,\frac{p+3q}{2}}(t)= \sum_{n=0}^{\infty}\sum_{{i,j=0}\atop i+j\leqslant n}^n
t^{-1-n}\frac{n!n!}{i!j!(n-i-j)!(j-q)!(i+\frac{q-p}{2})!
(n-i-j+\frac{p+q}{2})!}
\end{equation} 
for the triangular lattice. The latter 
can further be written as $H_{p'+q',2q'-p'}$, with $p'=q$ and
$q'=(p+q)/2$, by swapping $i$ and $j$ in the sum. 
Hence we only need to consider $H_{p+q,2q-p}(t)$, which can be written in
terms of gamma functions as
\begin{equation} 
H_{p+q,2q-p}(t) = \sum_{i,j,k=0}^{\infty} 
\frac{\G{1+i+j+k}^2}{i!j!k!\G{1-p+i}\G{1+p-q+j}\G{1+q+k}}\ t^{-1-i-j-k},
\end{equation}  
where the domain of the summing indices are extended to infinity taking 
into account the poles of gamma functions with non-positive arguments as
before. The series can be rewritten as 
\begin{equation}
H_{p+q,2q-p}(t)=\frac{1}{\G{1-p}\G{1+q}\G{1+p-q}} \frac{1}{t} 
F^{(3)}_C\left(
\left.
\genfrac{}{}{0pt}{}{1,1}{1-p,1+q,1+p-q}
\right| \frac{1}{t},\frac{1}{t},\frac{1}{t}\right),
\end{equation} 
for $|t|>3$, 
where $F^{(3)}_C$ denotes the third Lauricella function in three variables. 
While the expression is not in the domain of interest for the evaluation of
the Green's function, its values
near $t=1$ may be obtained by analytic continuation. 
The result for Kagome and dice lattices are obtained by the same $H$ at other
values of $t$ \cite{wu}.
\section{Higher dimensions}
The method generalizes to three and higher dimensions as well.
In the case of a  body-centered cubic lattice in $D$  dimensions, 
for example, with 
$\N = \{(\pm 1,\pm 1,\cdots, \pm 1)\}$ and $N=2^D$, the Green's function
\eq{green1} is  
\begin{equation} 
G^{\bcc}_{r_1,r_2,\cdots,r_D} =  
\left(\frac{1}{2\pi}\right)^D
\int_0^{2\pi}d\t_1 \cdots \int_0^{2\pi}d\t_D \
\frac{e^{ip\bcdot\t}}{2^D-2^D\prod\limits_{i=0}^D\cos\t_i}.
\end{equation} 
Defining angular variables as $x_i = e^{\t_i}$, 
for $i=1,2,\cdots D$ we write this as
\begin{equation} 
G^{\bcc}_{r_1,r_2,\cdots,r_D} =  H_{r_1,r_2,\cdots,r_D}(1),
\end{equation} 
with the resolvent  defined as a contour integral as before, namely,
\begin{equation} 
H_{r_1,\cdots, r_D}(t)= \left(\frac{1}{2\pi i}\right)^{D}
\oint\limits_{|x_1|=1}\frac{dx_1}{x_1}\cdots\oint\limits_{|x_D|=1} \frac{dx_D}{x_D}\
\frac{x_1^{r_1}\cdots x_D^{r_D}}
{2^Dt-\prod\limits_{i=0}^D (x_i+1/x_i)}.
\end{equation} 
Evaluating the contour integral in a similar manner as in the previous cases
through expansions using the geometric series and binomial series seriatim,
finally calculating the residue yields
\begin{equation}
\begin{split}
H_{r_1,\cdots, r_D}(t) &=  \frac{1}{\pi^D}\frac{1}{t}\prod_{i=1}^D
\frac{1}{r_i}\sin\left(\frac{\pi r_i}{2}\right)
{}_{2D+1}F_{2D}\left(\left.
\genfrac{}{}{0pt}{}{\overbrace{{1},\cdots,{1}}^{D+1},
\overbrace{1/2,\cdots,1/2}^{D}}%
{\frac{2+r_1}{2},\frac{2-r_1}{2},\cdots,\frac{2+r_D}{2},\frac{2-r_D}{2}}
\right|\frac{1}{t^2}\right) \\
&\quad +  \frac{1}{\pi^D}\frac{1}{t^2}\prod_{i=1}^D
\frac{1}{1-r^2_i}\cos\left(\frac{\pi r_i}{2}\right)
{}_{2D+1}F_{2D}\left(\left.
\genfrac{}{}{0pt}{}{\overbrace{1,\cdots,1}^{D+1},
\overbrace{3/2,\cdots,3/2}^{D}}%
{\frac{3+r_1}{2},\frac{3-r_1}{2},\cdots,\frac{3+r_D}{2},\frac{3-r_D}{2}}
\right|\frac{1}{t^2}\right)
\end{split}
\end{equation} 
in terms of the hypergeometric function.
\section{Inclusion of next-nearest neighbor hopping}
The scope of the method under consideration is even more general.  
It depends on writing cosines of
angular variables in terms of phases so as to interpret the integration over
each angular variable as a contour integration on a 
complex plane. This aspect generalizes to cases with longer range hopping as
well. Let us demonstrate this by evaluating the Green's function on  a
one-dimensional lattice with next-nearest neighbor hopping, namely,
\begin{equation}
G_r^{\nnn}(\tau_1,\tau_2) = 
\frac{1}{2\pi} \int_0^{2\pi} \frac{e^{ir\t}\ d\t}{%
2-2\tau_1\cos\t -2\tau_2\cos 2\t},%
\end{equation} 
where $\tau_1$ and $\tau_2$ are the hopping strengths of the nearest and the
next-nearest neighbors, respectively. Similar to the previous cases 
the Green's function can be written in terms of the contour integral 
\begin{equation} 
H_r^{\nnn}(\tau_1,\tau_2) = \frac{1}{2\pi i}
\oint\limits_{|x|=1} \frac{dx}{x} \frac{x^r}{%
\frac{2}{\tau_1} - (x+\frac{1}{x}) - \frac{\tau_2}{\tau_1} 
(x^2+\frac{1}{x^2})},
\end{equation} 
as $G_r^{\nnn}(\tau_1,\tau_2) = 
\frac{1}{\tau_1} H_r^{\nnn} ({\tau_1},{\tau_2})$. 
Following our method of evaluating the contour integral as the coefficient of
the constant term in a series expansion of the integrand  we obtain 
\begin{equation} 
\label{eq:nnn}
H_r^{\nnn}(\tau_1,\tau_2) = \sum\limits_{n,n',b,a=0}^{\infty}
\left(\frac{\tau_2}{\tau_1}\right)^{n'}\left(\frac{\tau_1}{2}\right)^{1+n} 
\frac{\G{1+n}}{\G{1+b}\G{1+a}\G{1+n'-b}\G{1+n-n'-a}},
\end{equation} 
with the  indices satisfying the constraint 
\begin{equation}
r+4b+2a-n'-n=0. 
\end{equation} 
Writing $n-n'-b=c$ and $n'-b=n''$ the constraint is solved for $n''$ as 
\begin{equation}
n'' = b  + \frac{r+a-c}{2}.
\end{equation} 
Substituting this in the series \eq{eq:nnn} and using the duplication formula
\eq{dup} in the numerator we obtain the expression for the resolvent as
\begin{equation} 
H_r^{\nnn}(\tau_1,\tau_2) = \frac{\tau_2^{r/2}\tau_1}{2\sqrt{\pi}}
\sum\limits_{a,b,c=0}^{\infty} \frac{
\left({\tau_1^2\tau_2}\right)^{\frac{a}{2}}
\left({\tau_2}\right)^{2b}
\left(\frac{\tau_2}{\tau_1^2}\right)^{\frac{c}{2}}
}{a!b!c!} \frac{
\G{1+\frac{3a+c+r}{4}+b}\G{\frac{1}{2}+\frac{3a+c+r}{4}+b}}{
\G{1+\frac{r+a-c}{2}+b}}.
\end{equation} 
This series can be expressed in terms of hypergeometric functions. In
particular, effecting the sum over $b$ it can be expressed as an infinite
series in the Gauss' hypergeometric function ${}_2F_1$. However, 
the particular manner of arranging the terms affects the domain of
convergence of the series, which in turn depends on the physical situation
under consideration. The Green's function for various limits of the
coupling parameters $\tau_1$ and $\tau_2$ can be studied using the series
by expressing it as Barnes' integral and deforming contours. 
\section{Discussion}
In this article we have studied Green's function on lattices as the kernel 
of the discrete Laplacian operator. We present a means to evaluate the 
resolvent 
exemplified by square lattices in any dimension as well as the triangular and
the honeycomb lattices in two dimensions. 
The Green's function, written as a Fourier transform is
first expressed in terms of a contour integral in as many variables as the
dimension of the lattice. The contour integral is  evaluated as the residue
of the integrand. The residue, defined as usual as the coefficient of the 
terms with inverses of the integration variables in the Laurent expansion 
of the integrand is obtained as a series in every case. We identify these
series as combination of generalized hypergeometric functions for the square
lattices. For the two-dimensional triangular and honeycomb lattices the
Greens' function is expressed in terms of the Lauricella function of the
third kind, which is a generalization of hypergeometric functions in three
variables. As a consequence of being able to express the Green's function in
terms of generalized hypergeometric functions, for the two-dimensional square
lattice we exhibit the cancellation of
the singularity of the self-energy term, which is subtracted from the Green's
function to impose translational symmetry, using existing results on
analytic continuation of zero-balanced generalized hypergeometric functions.
We show that depending on the way of arranging terms of the series
the same Green's function may be expressed as different combinations
hypergeometric functions. This provide identities between hypergeometric
functions. Physically, moreover, it corresponds to respecting or
breaking the symmetries of the lattice by the Green's function. The form
obtained by breaking the isotropy of the lattice is better-suited for
studying the asymptotics, namely, the behavior of the Green's function at
large distances \cite{martin,scaling}. 
Since the resolvents are written in terms of hypergeometric functions in all
the cases considered here, it is  possible to write them as solutions to
certain Fuchsian equations in the spectral parameter using the dfferential
equations satisfied by the corresponding hypergeometric functions. 
It will be interesting to relate these to geometry by interpreting the 
contour integrals as periods of appropriate algebraic varieties 
\cite{senray,guttmann} and the Fuchsian equation as the equation for
existence of flat connection in the moduli space. 

We expect the method presented here to be of use in various other cases as
well. 
\section*{Acknowledgement}
It is a pleasure to thank Pushan Majumdar, Avijit Mukherjee and Krishnendu
Sengupta for illuminating discussions.

\end{document}